# Ultrafast Carrier Dynamics in Multiple Quantum Well P-I-N Photodiodes

Yifan Zhao[1,2] and Mona Jarrahi[1,2,*]

[1] Electrical and Computer Engineering Department, University of California, Los Angeles, CA, 90095, USA

[2] California NanoSystems Institute, University of California, Los Angeles, CA, 90095, USA

*mjarrahi@ucla.edu

**Abstract**

Quantum well (QW) structures are widely used in lasers, semiconductor optical amplifiers, modulators, enabling their monolithic integration on the same substrate. As optoelectronic systems evolve to meet the growing bandwidth demands in the terahertz regime, a deep understanding of ultrafast carrier dynamics in QW structures becomes essential. We introduce a comprehensive model to analyze the ultrafast dynamics of interband photo-excited carriers in QW p-i-n structures and to calculate their frequency response. This model characterizes the entire photocarrier transport process, including carrier escape from QWs and movement across heterojunction interfaces. Additionally, we outline theoretical methods for calculating carrier escape times from both QWs and heterojunction interfaces. Using a GaAs/AlGaAs QW p-i-n structure as a case study, we discuss the effects of carrier escape times from QWs and heterojunction interfaces, as well as carrier transit time through the intrinsic region, on the frequency response of QW p-i-n structures.

## I. Introduction

Quantum wells (QWs) are engineered semiconductor structures composed of thin layers that exhibit quantum effects [1-3]. These structures have found widespread application in lasers [4-6] and semiconductor optical amplifiers (SOAs) [7-9], owing to their low threshold current, high quantum efficiency, and wavelength tunability. Another significant application is in QW-based electroabsorption and phase modulators [10-12], which are valued for their high modulation efficiency and high-speed modulation [13-16]. Furthermore, QW structures enable monolithic integration of various optical components, including lasers, SOAs, modulators, and photodiodes [17-21].

      One of the most widely used QW wafer configurations is the p-i-n structure, where an intrinsic region containing multiple quantum wells (MQWs) is sandwiched between n-type and p-type cladding layers. It is well established that the speed of devices based on p-i-n structures is limited by the carrier transit time and the circuit's resistance-capacitance (RC) time constant. However, a thorough analysis of carrier dynamics and speed limitations in QW p-i-n structures is still lacking. Previous studies have

primarily focused on the impacts of carrier transit time and the RC time constant [18, 22-29]. It has been suggested that the time required for excited carriers to escape from the QWs should also be considered [30, 31]. Nonetheless, the effect of carrier escape time on the overall response time of QW p-i-n structures has mostly been addressed intuitively, with no physical models available for quantitative analysis [18, 23, 27]. As optoelectronic systems continue to evolve toward the terahertz regime, a comprehensive understanding of ultrafast carrier dynamics in QW p-i-n structures and precise derivation of their frequency response are increasingly important.

In this paper, we present a comprehensive model to analyze the ultrafast dynamics of interband photo-excited carriers in QW p-i-n structures. Our model characterizes the entire photocarrier transport process, including the carrier escape process from QWs and heterojunction interfaces within the depletion region. Additionally, we provide theoretical methods for calculating carrier escape times from both QWs and heterojunction interfaces. A general framework for calculating the frequency response of QW p-i-n heterostructures is introduced and applied to a case study of GaAs/AlGaAs QW p-i-n heterostructures. Finally, we discuss the impact of carrier escape times from QWs and heterojunction interfaces, along with carrier transit time through the intrinsic region, on the frequency response of QW p-i-n heterostructures.

## II. Frequency response of QW p-i-n heterostructures

In this section, we present a theoretical model to analyze ultrafast carrier dynamics in QW structures. For generality, we consider a p-i-n QW structure as depicted in Fig. 1. This structure consists of a thin intrinsic region of thickness $L$ with multiple QW pairs sandwiched between p-type and n-type cladding layers, forming heterojunctions at the interfaces. When photons are absorbed within the QWs, photocarriers are generated. These photocarriers escape from the QWs, drift across the intrinsic region, and may become temporarily trapped at the heterojunction interface due to an energy barrier. Eventually, the carriers overcome this barrier, enter and traverse the cladding layers, and are collected by the p- and n-doped contact layers.

We assume that the escape processes from the QW and heterojunction interfaces can each be described by exponential decay, with time constants $\tau_{QW,e(h)}$ and $\tau_{HJ,e(h)}$ for electrons (holes), respectively. Additionally, we assume that the electric field is strong enough to accelerate the escaped carriers to their saturation velocities immediately. Carrier recombination within the intrinsic region is ignored, as the carrier lifetime for QW structures is typically on the order of tens of nanoseconds at room temperature [32], which is much longer than the time scales of the carrier transit and trapping processes associated with the QW p-i-n heterostructures studied here. Under high reverse bias voltages, we can also neglect the probability of escaped carriers being re-trapped by adjacent QWs [33]. To derive the frequency response, we start with the time-domain impulse response and apply a Fourier transform. For simplicity, we express the response for photo-generated electrons in the *m*-th well, as similar expressions apply for holes and for carriers generated in other QWs, allowing us to linearly sum the results.

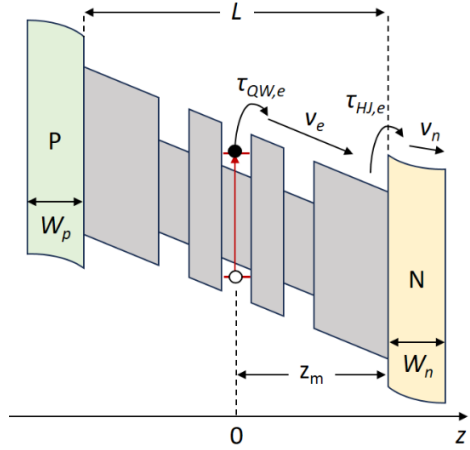

**Figure 1.** Illustration of the dynamics of a photo-generated electron in a QW.

At time $t = 0$, consider an optical impulse incident on the structure, generating an electron-hole pair density $N_m$ in the *m*-th QW. The 2D electron density within the *m*-th well is given by:

$$N(t) = N_m e^{-t/\tau_{QW,e}} u(t) \tag{1}$$

where $u(t)$ is the step function. Assuming electrons move in $+z$ direction, the electron density in the intrinsic region is:

$$n(z,t) = \frac{N_m}{v_e \tau_{QW,e}} e^{-\left(t-\frac{z}{v_e}\right)/\tau_{QW,e}} u\left(t - \frac{z}{v_e}\right) \tag{2}$$

with $z = 0$ at the center of the *m*-th well and $v_e$ as the electron saturation velocity. According to the Shockley-Ramo theorem [34-36], the external circuit current induced by electrons moving in the intrinsic region, originating from the *m*-th well, is:

$$I_{e,m,i}(t) = f_i \int_0^{z_m} q v_e S n(z,t) \, dx = \begin{cases} f_i q v_e S N_m \left(1 - e^{-\frac{t}{\tau_{QW,e}}}\right), & 0 < t < t_m \\ f_i q v_e S N_m \left(1 - e^{-\frac{t_m}{\tau_{QW,e}}}\right) e^{-\frac{t-t_m}{\tau_{QW,e}}}, & t \geq t_m \end{cases} \tag{3}$$

where $S$ is the device area, $f_i = \frac{1/\varepsilon_i}{L/\varepsilon_i + W_n/\varepsilon_n + W_p/\varepsilon_p}$ is the weighting field in the intrinsic region, $\varepsilon_i$, $\varepsilon_n$ and $\varepsilon_p$ are the dielectric constants of the intrinsic, n-cladding, and p-cladding layers, $W_n$ and $W_p$ are the thicknesses of the depleted n- and p-cladding layers, respectively; $z_m$ is the distance from the center of the *m*-th well to the n-side heterojunction interface; and $t_m = \frac{z_m}{v_e}$ is the electron transit time within the intrinsic region. The frequency-domain expression is obtained by Fourier transforming:

$$I_{e,m,i}(\omega) = \int I_{e,m,i}(t)e^{-j\omega t}dt = z_m f_i \frac{qSN_m}{1+j\omega\tau_{QW,e}} \text{sinc}\left(\frac{\omega t_m}{2}\right)e^{-\frac{j\omega t_m}{2}} \quad (4)$$

The rate equation for electrons trapped at the heterojunction interface is

$$\frac{dN_t}{dt} = -\frac{N_t}{\tau_{HJ,e}} + n(z_m,t)v_e \quad (5)$$

where $N_t$ is the 2D density of trapped electrons, and $n(x_m,t)$ is the electron density at the interface. The solution to this equation is:

$$N_t(t) = \frac{N_m}{1-\frac{\tau_{QW,e}}{\tau_{HJ,e}}}\left(e^{-\frac{t-t_m}{\tau_{HJ,e}}} - e^{-\frac{t-t_m}{\tau_{QW,e}}}\right)u(t-t_m) \quad (6)$$

The external circuit current induced by electrons moving in the depleted n-cladding layer, originating from the $m$-th well, is given by:

$$I_{e,m,n}(t) = f_n q v_n S \int_{t-t_n}^{t} \frac{N_t(t)}{\tau_{HJ,e}}dt \quad (7)$$

where $t_n = \frac{W_n}{v_n}$ is the electron transit time in the depleted n-cladding layer, $v_n$ is the saturation velocity, and $f_n = \frac{1/\varepsilon_n}{L/\varepsilon_i + W_n/\varepsilon_n + W_p/\varepsilon_p}$ is the weighting field of the n-cladding layer. The frequency-domain expression is:

$$I_{e,m,n}(\omega) = W_n f_n \frac{qSN_m}{1+j\omega\tau_{QW,e}} \frac{1}{1+j\omega\tau_{HJ,e}} \text{sinc}\left(\frac{\omega t_n}{2}\right)e^{-j\omega\left(t_m+\frac{t_n}{2}\right)} \quad (8)$$

To obtain the total external circuit current, we sum the contributions from both electrons and holes generated in all QWs. These components can be derived by incorporating the appropriate weighting fields, lengths, and time constants into Eqs. 4 and 8. The overall frequency response due to carrier dynamics in the QW and heterojunction structures is given by:

$$H(\omega) = \frac{1}{\sum_m qSN_m}\sum_m\left[I_{e,m,i}(\omega) + I_{e,m,n}(\omega) + I_{h,m,i}(\omega) + I_{h,m,p}(\omega)\right] \quad (9)$$

where $I_{h,m,i}$ and $I_{h,m,p}$ represent the external circuit currents induced by holes originated from the $m$-th well and moving in the intrinsic and p-cladding layers, respectively.

## III. Carrier escape time from QWs

In this section, we outline a method for calculating the escape time constant for electrons and holes in a QW structure. We consider a symmetric QW with a barrier height $E_b$ and width $w$, subject to an electric field $F$ along the -z direction, as shown in Fig. 2. Within the well region ($-w < z < 0$), the conduction band profile is given by $V(z) = -Fz$, while outside the well ($z < -w$ or $z > 0$), the profile is $V(z) = E_b - Fz$. Under the influence of the electric field, electron states with energy below the barrier become quasi-bounded, allowing for tunneling with a finite probability. The combined tunneling and thermionic emission currents form the total electron escape current, with the escape time constant defined as $\tau_{QW,e} = \frac{N_{QW,e}}{J_e}$, where $J_e$ is the electron escape current density and $N_{QW,e}$ is the 2D electron density in the well.

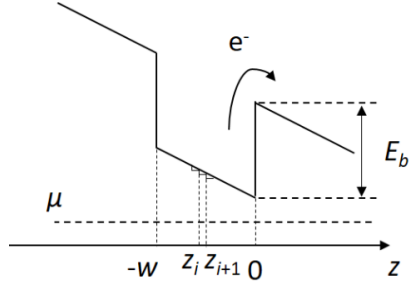

**Figure 2.** The conduction band profile of a QW with barrier height $E_b$ and width $w$ under electric field along z-axis.

To calculate $J_e$, we first consider the electron state in k-space, represented as $\frac{dk_x dk_y dk_z}{(2\pi)^3}$, for the wave vectors $k_x$, $k_y$, and $k_z$ in each direction [37]. The escape current for a single electron in that state is given by $q\langle v_g^+(E_z)\rangle T^+(E_z)$, where $\langle v_g^+(E_z)\rangle$ is the average group velocity in the +z direction and $T^+(E_z)$ is the transmission probability at the right barrier. The total escape current is calculated by integrating over all states weighted by the Fermi distribution [38, 39]:

$$J_e = q \iint_{-\infty}^{\infty} \frac{dk_x dk_y}{(2\pi)^2} \left[ \int_{k_{th}}^{\infty} f(E)\langle v_g^+(E_z)\rangle T^+(E_z) \frac{dk_z}{2\pi} \right] \quad (10)$$

where $f(E) = \frac{1}{e^{(E-\mu)/kT}+1}$ is the Fermi-Dirac distribution with Fermi level $\mu$, $E = E_t + E_z$ is the total energy decomposed into in-plane ($E_t$) and perpendicular ($E_z$) components, and $k_{th}$ denotes a threshold below which the tunneling is negligible, taken as the energy where $T^+(E_{th}) = 10^{-3}$. Reformulating in terms of energy, we obtain:

$$J_e = q \int_0^{\infty} \frac{m_e}{2\pi\hbar^2} dE_t \left[ \int_{E_{th}}^{\infty} f(E)\langle g_{1d}(E_z)\rangle \langle v_g^+(E_z)\rangle T^+(E_z) dE_z \right] \quad (11)$$

where $m_e$ is the electron effective mass, $\langle g_{1d}(E_z)\rangle$ represents the average 1D density of states (DOS), and $E_{th} = E_z(k_{th})$. Below the threshold, we consider the states to be well-bounded with discrete, step-like DOS, similar to un-biased QWs. Above the threshold, states leak out of the QW and are broadened with continuous DOS, contributing to the escape current. It should be noted that a lower tunneling threshold level does not change the estimated current value considerably.

To determine $\langle g_{1d}(E_z)\rangle$, $\langle v_g^+(E_z)\rangle$ and $T^+(E_z)$, we re-write the Schrödinger equation as

$$\frac{dG(z, E_z)}{dz} = -j\left[\frac{m_e(z)}{2\hbar} G^2(z, E_z) + \frac{4}{\hbar}[V(z) - E_z]\right] \quad (12)$$

where $G(z, E_z) = \frac{2\hbar}{jm_e(z)}\left[\frac{1}{\psi(z)}\frac{d\psi(z)}{dz}\right]$ with $\psi(z)$ representing the electron wavefunction and $V(z)$ is the potential, which is a function of the applied bias. We define the energy reference at $z = 0$ inside the well. As shown in [40], $G(z, E_z)$ can be computed analogously to impedance in transmission lines. For the z-axis discretized into differential segments at positions $z_i$, the transmission line impedance seen from +z and –z directions, $G^+$ and $G^-$, is calculated iteratively using

$$G^+(z_i) = G_0(z_i)\frac{G^+(z_{i+1})\cosh(\xi_i \Delta z_i) - G_0(z_i)\sinh(\xi_i \Delta z_i)}{G_0(z_i)\cosh(\xi_i \Delta z_i) - G^+(z_{i+1})\sinh(\xi_i \Delta z_i)} \quad (13)$$

$$G^-(z_{i+1}) = G_0(z_i)\frac{G^-(z_i)\cosh(\xi_i \Delta z_i) + G_0(z_i)\sinh(\xi_i \Delta z_i)}{G_0(z_i)\cosh(\xi_i \Delta z_i) + G^-(z_i)\sinh(\xi_i \Delta z_i)} \quad (14)$$

where $\Delta z_i = z_{i+1} - z_i$ is the length of the i-th segment, $G_0(z_i) = 2\hbar \xi_i/jm_e(z_i)$ is the characteristic impedance of the i-th segment, and $\xi_i = j\sqrt{2m_e(z_i)(E_z - V(z_i))}/\hbar$ is the propagation constant along the i-th segment. From the transmission line analogy, the transmission probability $T^+(E_z)$ at the QW boundary is then determined by $T^+(E_z) = 1 - |r^+|^2$, where $r^+ = \frac{G^+(0) - G_0(0)}{G^+(0) + G_0(0)}$ is the reflection coefficient.

For electrons with $E_z < E_{th}$, the energy levels can be solved from the condition that $G^+(z, E_l) = G^-(z, E_l)$ at any position, where $E_l$ is the energy of l-th well-bounded state [40]. For electrons with $E_z > E_{th}$, the DOS becomes continuous and the 1D DOS (including both spins) can be calculated as $g_{1D}(z, E_z) = Im\left[\frac{j8/\pi\hbar}{G^+(z,E_z) - G^-(z,E_z)}\right]$ while the group velocities of the electrons moving in the +z and –z directions are given by $v_g^\pm(z, E_z) = \frac{1}{2}Re[G^\pm(z, E_z)]$ [39]. Averaging over the QW regions gives the values $\langle g_{1d}(E_z)\rangle$ and $\langle v_g^+(E_z)\rangle$.

$$\langle g_{1D}(E_z)\rangle = \frac{1}{w}\int_{-w}^{0} g_{1D}(z, E_z)dz \quad (15)$$

$$\langle v_g^+(E_z)\rangle = \frac{1}{w\langle g_{1D}(E_z)\rangle}\int_{-w}^{0} v_g^+(z, E_z)g_{1D}(z, E_z)dz \quad (16)$$

The 2D electron density $N_{QW,e}$ is then computed as:

$$N_{QW,e} = w\int_{0}^{\infty} g_{3d}(E)f(E)dE \quad (17)$$

where the 3D DOS $g_{3d}(E)$ varies with energy. For electrons with $E < E_{th}$, the step-like DOS for discrete states is expressed by [41]:

$$g_{3d}(E)\big|_{E<E_{th}} = \sum_{E_l<E}\frac{m_e}{\pi^2\hbar^2}(k_l - k_{l-1}) \quad (18)$$

where $k_l = \sqrt{2m_e E_l}/\hbar$ and $k_0 = 0$. For electrons with $E > E_{th}$, the continuous DOS is derived by convolving the 1D and 2D in-plane DOS given by $\frac{m_e}{2\pi\hbar^2}$ [2]:

$$g_{3d}(E) = \sum_{\substack{E>E_{th} \\ E_l<E_{th}}} \frac{m_e}{\pi^2\hbar^2}(k_l - k_{l-1}) + \int_0^\infty \frac{m_e}{2\pi\hbar^2} dE_t \int_{E_{th}}^{E} \langle g_{1d}(E_z)\rangle \delta(E - E_t - E_z) dE_z \quad (19)$$

For holes, we assume that heavy and light holes are in equilibrium with a common quasi-Fermi level. We calculate the escape current and carrier density for heavy and light holes separately, then we sum them up for the total current and hole density. Whereas the bulk effective mass is used for states with $E > E_{th}$, in-plane masses calculated from k·p method for different sub-bands are used for $E < E_{th}$ states.

## IV. Carrier trapping time at the heterojunction interface

This section outlines our method for calculating carrier trapping time at a heterojunction interface due to the presence of an energy barrier. Consider a heterojunction with a conduction band barrier $E_b$ as shown in Fig. 3. The left side is intrinsic, and the right side is n-doped with a doping level $N_2$, $\mu_1$ and $\mu_2$ represent the Fermi levels on the left and right sides, respectively. Under an electric field $F$, the conduction band diagram is described by $E_c(z) = -Fz$ for $z < 0$ and $E_c(z) = E_b + \frac{1}{2}FW_n[1 - (1 - \frac{z}{W_n})^2]$ for $0 < z < W_n$, where $W_n = \frac{\varepsilon_2\varepsilon_0 F}{qN_2}$ is the depletion region width on the n-doped side.

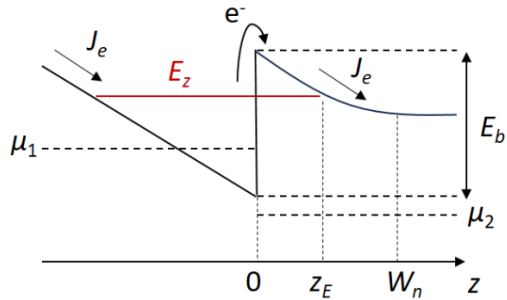

**Figure 3.** Conduction band profile at the interface between the intrinsic layer and n-cladding layer.

In a steady state, electrons are injected from the left side with a current density $J_e$, and the same magnitude of current is emitted over the heterojunction barrier to the right side. Carriers trapped at the heterojunction interface can overcome the barrier either by thermionic emission or tunneling. The electron transmission probability through the barrier is given by [42]:

$$T(E_z) = \begin{cases} \exp\left(-\frac{2}{\hbar}\int_0^{z_E}[2m_{e2}(E_c(z) - E_z)]^{1/2} dz\right), & \text{if } E_{min} \leq E_z \leq E_c(0^+) \\ 1, & \text{if } E_z \geq E_c(0^+) \end{cases} \quad (20)$$

where $E_c(z)$ is the conduction band profile, $E_{min} = \max(E_c(0^-), E_c(W_n))$ is the minimum energy level for electron tunneling, $z = 0$ is the position of the heterojunction interface, $z_E$ is the position where $E_c(z_E) = E_z$, and $m_{e2}$ is the electron effective mass in the cladding layer.

The total current, which includes both thermionic emission and tunneling components, is calculated as

$$J_e = J_{tunn} + J_{therm} = \frac{AT}{k}\int_{E_{min}}^{E_c(0^+)} f'(E_z)T(E_z)\,dE_z + \frac{AT}{k}\int_{E_c(0^+)}^{\infty} f'(E_z)T(E_z)\,dE_z \quad (21)$$

where $A = \frac{4\pi q m_{e1} k^2}{h^3}$ is the Richardson constant, $m_{e1}$ is the effective electron mass on the intrinsic side, $k$ is the Boltzmann constant, and $f'(E_z) = \ln(1 + \exp[(\mu_1 - E_z)/kT])$ is the Fermi distribution integrated over in-plane states. The current in the opposite direction is negligible under high electric fields.

The 2D density of trapped electrons, $N_{HJ,e}$, is calculated by integrating the electron density over the intrinsic region:

$$N_{HJ,e} = \int_0^{-\infty} N_c e^{-(E_c(z)-\mu_1)/kT}\,dz \quad (22)$$

where $N_c$ is the effective conduction band density of states in the intrinsic region. The electron trapping time at the heterojunction interface is given by $\tau_{HJ,e} = \frac{N_{HJ,e}}{J_e}$.

The hole trapping time at the heterojunction interface, $\tau_{HJ,h} = \frac{N_{HJ,h}}{J_h}$, is calculated similarly. We derive it by using the effective valence band density of states in the intrinsic region, $N_v$, and the density of state mass $m_{dos} = (m_{hh}^{\frac{3}{2}} + m_{lh}^{\frac{3}{2}})^{2/3}$, where $m_{hh}$ and $m_{lh}$ are the effective masses of heavy and light holes, respectively [42].

In certain materials, such as GaAs/Al$_x$Ga$_{1-x}$As ternary systems, additional considerations are needed since Al$_x$Ga$_{1-x}$As becomes indirect bandgap for $x > 0.45$. In these cases, the electron transport across the barrier involves Γ-X intervalley scattering, which can be incorporated by a reduced effective Richardson constant, experimentally characterized in such systems [43]. It has been shown that tunneling current across the GaAs/AlGaAs heterojunction for indirect bandgap AlGaAs deviates from predictions by a prefactor, displaying similar reduction trends to the effective Richardson constant [43]. In our model we use the same effective Richardson constant to account for the inter-valley scattering effect for both thermionic emission and tunneling. In addition, we use the effective mass of the final state after tunneling for the tunneling probability calculations, as will be detailed in later sections.

## V. Discussion

We applied our model to a commonly used GaAs/AlGaAs QW gain heterostructures (Table 1), which features a p-i-n structure with an ~110 nm Al$_{0.3}$Ga$_{0.7}$As intrinsic region, comprising three pairs of Al$_{0.08}$

Ga$_{0.92}$As/Al$_{0.3}$Ga$_{0.7}$As QWs, situated between two Al$_{0.55}$Ga$_{0.45}$As cladding layers. The Fermi level at each applied electric field is derived from the experimentally measured current density, $J_e$. Figure 4a illustrates the calculated electron and hole escape times from the QWs for electric fields ranging from 10 to 300 kV/cm. For both carriers, escape times decrease exponentially as the electric field increases due to the narrowing of the barrier width. At relatively low electric fields (below approximately 200 kV/cm), our results show that holes have a shorter escape time than electrons. However, at higher electric fields, electrons exhibit shorter escape times. This difference arises because holes have a lower barrier height and a greater number of accessible energy levels due to their higher effective mass. At high electric fields, however, the barrier shape becomes triangular, causing the effective barrier width to shrink rapidly. As a result, tunneling becomes more efficient for electrons, given their smaller effective mass. At an electric field of 300 kV/cm (corresponding to approximately -3V bias on the QW structure), the escape times for both electrons and holes reach approximately 0.1 ps. This indicates the potential of QW structures for high-speed photodiode applications, supporting frequencies above 100 GHz and even in the terahertz range.

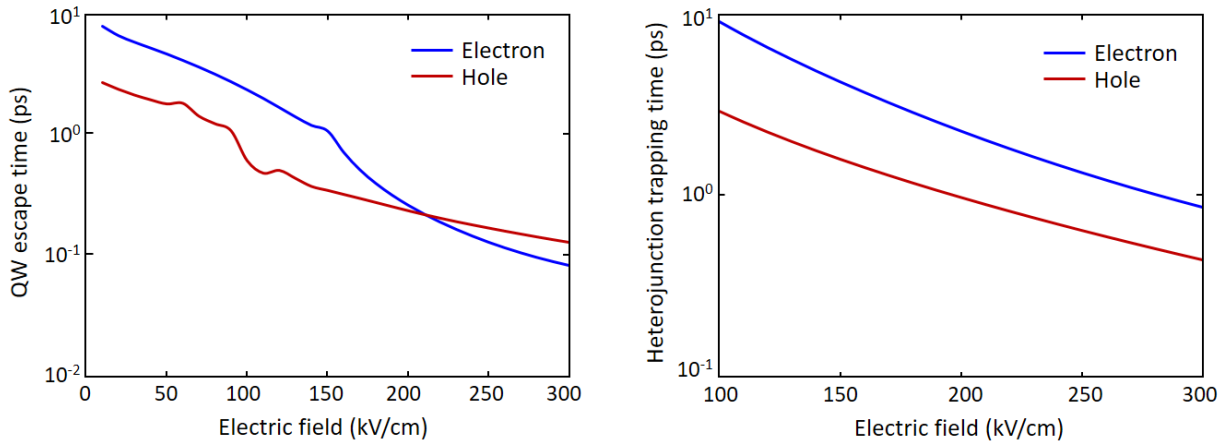

**Figure 4.** The calculated electron and hole escape times from the QWs (left) and trapping times at the heterojunction interfaces (right) for the QW p-i-n structure shown in Table 1.

Our calculated heterojunction trapping times for the QW structure described in the previous section are presented in Figure 4b as a function of the electric field applied to the substrate. The heterojunction trapping time decreases exponentially with increasing electric field due to the narrowing of the barrier width. The parameters used for these calculations are provided in Table 2. The Fermi level at each electric field was derived from the experimentally measured current density, $J_e$. Notably, the heterojunction trapping time is approximately 5–10 times longer than the QW escape time.

| Material | Mole Fraction (x) | Thickness (nm) | Type | Doping level (cm$^{-3}$) | Description |
|---|---|---|---|---|---|
| GaAs | | 200 | p-doped(C) | >2×10$^{19}$ | P-contact |
| Al(x)GaAs | 0.55 to 0.05 | 50 | p-doped(C) | >3×10$^{18}$ | |
| Al(x)GaAs | 0.55 | 1250 | p-doped(C) | 1×10$^{18}$ | P-cladding |
| GaIn(x)P | 0.49 | 10 | p-doped(Zn) | 7×10$^{17}$ | Etch stop |
| Al(x)GaAs | 0.55 | 300 | p-doped(C) | 6×10$^{17}$ | P-cladding |
| Al(x)GaAs | 0.3 | 40 | Undoped | | Barrier |
| Al(x)GaAs (×2) | 0.08 | 5.5 | Undoped | | QW |
| Al(x)GaAs (×2) | 0.3 | 6 | Undoped | | Barrier |
| Al(x)GaAs | 0.08 | 5.5 | Undoped | | QW |
| Al(x)GaAs | 0.3 | 40 | Undoped | | Barrier |
| Al(x)GaAs | 0.55 | 1500 | n-doped(Si) | 1×10$^{18}$ | N-cladding |
| Al(x)GaAs | 0.05 to 0.55 | 50 | n-doped(Si) | 2×10$^{18}$ | N-cladding |
| GaAs | | 1000 | n-doped(Si) | 2×10$^{18}$ | N-contact |
| GaAs | | | Undoped | | Substrate |

**Table 1.** The studied GaAs/AlGaAs QW heterostructure.

| | |
|---|---|
| Γ-Valley band gap $E_\Gamma$ (eV) | 1.424+1.247$x$ for $x$ < 0.45 |
| | 1.424+1.247$x$+1.147×($x$-0.45)$^2$ for $x$ < 0.45 |
| X-Valley band gap $E_X$ (eV) | 1.9+0.125$x$+0.143$x^2$ |
| Band split ratio | 65/35 |
| Electron effective mass (Γ-Valley) $m_{e\Gamma}$ ($m_0$) | 0.067+0.083$x$ |
| Electron effective mass (X-Valley transverse) $m_{eXt}$ ($m_0$) | 0.23-0.04$x$ |
| Heavy hole effective mass $m_{hh\perp}$ ($m_0$) | 0.62+0.14$x$ |
| Heavy hole effective mass $m_{hh\parallel}$ (in-plane) ($m_0$) | 0.734, 0.284, 0.454, 0.415 |
| Light hole effective mass $m_{lh\perp}$ ($m_0$) | 0.087+0.063$x$ |
| Light hole effective mass $m_{hh\parallel}$ (in-plane) ($m_0$) | 0.596, 0.222 |
| Hole effective mass (DOS) ($m_0$) | 0.51+0.25$x$ |
| Electron effective Richardson constant $A_e$ (A/cm$^2$K$^2$) | 1 |
| Hole Richardson constant $A_h$ (A/cm$^2$K$^2$) | 70 |
| Dielectric constant $\varepsilon_r$ | 13.18-3.12$x$ |
| Effective conduction band density of states $N_c$ ($x$<0.45) (cm$^{-3}$) | 2.5×10$^{19}$×(0.067+0.083$x$)$^{3/2}$ |
| Effective valence band density of states $N_v$ (cm$^{-3}$) | 2.5×10$^{19}$×(0.51+0.25$x$)$^{3/2}$ |
| Electron saturation velocity (10$^7$ cm/s$^{-1}$) | 0.72($x$=0.3), 0.58($x$=0.55) |
| Hole saturation velocity (10$^7$ cm/s$^{-1}$) | 0.8($x$=0.3), 0.8($x$=0.55) |
| QW escape time at 280 kV/cm (ps) | 0.0917(electrons), 0.135(holes) |
| Heterojunction trap time at 280 kV/cm (ps) | 0.98(electrons), 0.48(holes) |

**Table 2.** Al$_x$Ga$_{1-x}$As parameters used in our calculations.

It is important to note that the structure considered here has an indirect bandgap $Al_{0.55}Ga_{0.45}As$ cladding layer, which affects the escape dynamics: while electrons undergo intervalley scattering, holes do not. Previous studies [43] have shown that in indirect bandgap AlGaAs, intervalley transport across the barrier occurs from the Γ point to the transverse X bands, rather than the longitudinal X band. This is due to the smaller transverse effective mass ($0.21m_0$) of the X valley compared to its longitudinal effective mass ($1.19m_0$) [44], where $m_0$ is the free electron mass. This scattering process reduces the electron escape current via two mechanisms. First, the intervalley scattering involving phonons for momentum conservation is relatively inefficient, which lowers the effective Richardson constant. Second, the electrons in the X-valley have a larger effective mass than those in the Γ-valley, making tunneling events less likely. These factors are reflected in our results, which show that electrons exhibit longer trapping times than holes across the entire range of applied electric fields.

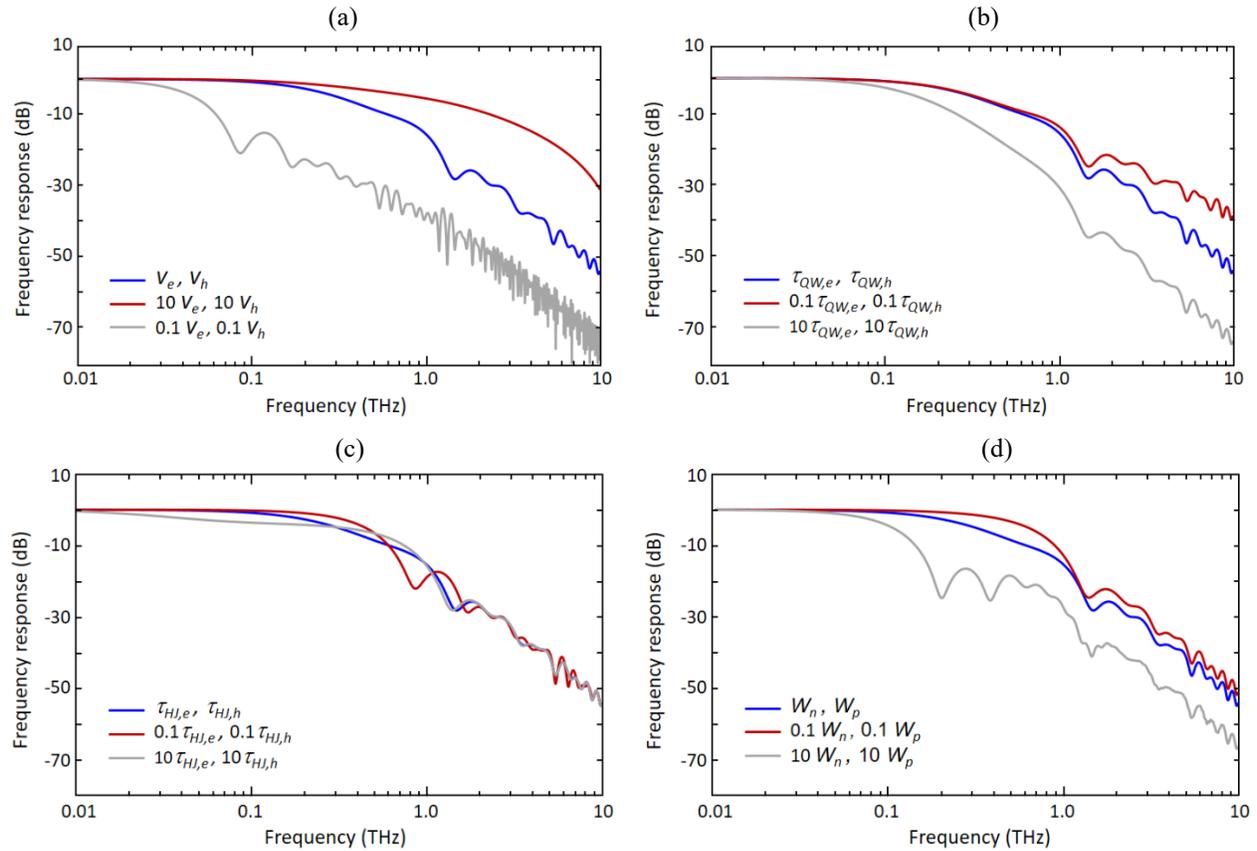

**Figure 5.** Frequency response of the QW p-i-n structure described in Table 1 is shown in blue. The frequency response when scaling the carrier transit time ($\tau_{tr}$), QW escape time ($\tau_{QW}$), heterojunction trapping time ($\tau_{HJ}$), and depletion width of the cladding layers ($W_n$ and $W_p$) by factors of 0.1 and 10 are shown in (a), (b), (c), and (d), respectively.

Having calculated the carrier escape times from both QWs and heterojunction interfaces, we examine the frequency response of the QW p-i-n structure shown in Table 1 to an optical excitation. Since this study does not involve a specific device structure, we exclude the impact of the RC response and focus solely on carrier dynamics. We assume that carriers are generated equally across the QWs and account for the varying distances from each QW to the heterojunction, summing their contributions using Eq. 9.

The blue plot in Fig. 5 illustrates the frequency response based on the calculated QW escape time and heterojunction trapping time at an applied field of 280 kV/cm (values provided in Table 2). For frequencies between 0.1 and 1 THz, the frequency response exhibits a 20 dB/decade slope due to the slow transit time of ~1 ps. At higher frequencies, the slope increases to 40 dB/decade as the QW escape time (~0.1 ps) introduces an additional bottleneck. Unlike classical expressions used in the literature, the transit time affects the frequency response via a sinc function rather than the $(1 + j\omega\tau_{tr})^{-1}$ form considered in other studies. This is clearly illustrated in Fig. 5a, where we vary all transit times by factors of 0.1 and 10. However, the QW escape time affects the frequency response with a $(1 + j\omega\tau_{QW})^{-1}$ dependence as illustrated in Fig. 5b, where the same scaling factors of 0.1 and 10 are applied to the QW escape times.

Phase difference between carriers generated in different QWs leads to an interference, which appears as ringing in the frequency response. Figure 5c compares frequency responses for various heterojunction trapping times, where we similarly scale the calculated heterojunction trapping times by factors of 0.1 and 1. Notably, the heterojunction trapping time does not significantly impact the overall frequency response envelope. This is because the heterojunction trapping time only affects the current contribution from the photocarriers moving inside the depleted n- and p-cladding layers with depletion widths $W_n$ and $W_p$, respectively. In the structure studied, the doping level is high enough to keep the depletion width in the cladding layer relatively small compared to the intrinsic region, resulting in minimal effect on the frequency response. This changes when the cladding layer is lightly doped, causing the depletion width to approach the width of the intrinsic region. This scenario is illustrated in Fig. 5d, where we vary the depletion width of the cladding layers by factors of 0.1 and 10.

In summary, we present a comprehensive model for analyzing the ultrafast dynamics of photo-generated carriers in p-i-n QW structures and for calculating their frequency response. Theoretical models for determining carrier escape time from QWs and trapping time at heterojunction interfaces are also provided. Using a GaAs/AlGaAs MQW gain substrate, commonly employed in conventional lasers and SOAs, we demonstrate that carrier trapping within QWs and at heterostructure interfaces does not impede high-speed photodiode performance at frequencies exceeding 100 GHz, extending even into the terahertz range. With careful optimization of heterostructure layer thickness, doping levels, and photodiode geometry, sub-picosecond carrier transit times and RC time constants can be achieved, facilitating the monolithic integration of terahertz optoelectronic devices and systems on a single MQW p-i-n chip.

**Acknowledgements**

The authors gratefully acknowledge the financial support from the Office of Naval Research (grant # N000142212531) and IET Harvey Engineering Research Prize. The studies on interband photomixing and ultrafast carrier dynamics in quantum well heterostructures were supported by the Department of Energy (grant # DE-SC0016925).